\documentclass{elsart3p}

\usepackage{amsmath,amssymb,epsfig}
\usepackage{units}
\def\fly{\ifmmode Y_{\kern -1.5pt\textrm{\scriptsize fl}}\else
                 $Y_{\kern -1.5pt\textrm{\scriptsize fl}}$\fi}
\def\mind #1_#2{%
   \ifmmode #1_{\kern -0.5pt\textrm{\scriptsize #2}}\else
           $#1_{\kern -0.5pt\textrm{\scriptsize #2}}$\fi}

\begin{document}

\begin{frontmatter}



\title{A novel method for the absolute fluorescence yield measurement \\ by  AIRFLY}

% \address[label2]{ 
\author[label2]{{\bf AIRFLY Collaboration}:\quad M. Ave},
\author[label3]{M. Bohacova\corauthref{cor1}},\ead{bohacova@fzu.cz}
\author[label4]{B. Buonomo},
\author[label2]{N. Busca},
\author[label2]{L. Cazon},
\author[label5]{S.D. Chemerisov},
\author[label5]{M.E. Conde},
\author[label5]{R.A. Crowell},
\author[label6]{P. Di Carlo},
\author[label7]{C. Di Giulio},
\author[label8]{M. Doubrava},
\author[label4]{A. Esposito},
\author[label9]{P. Facal},
\author[label5]{F.J. Franchini},
\author[label1]{J.R. H\"orandel},
\author[label3]{M. Hrabovsky},
\author[label6]{M. Iarlori},
\author[label5]{T.E. Kasprzyk},
\author[label1]{B. Keilhauer},
\author[label10]{H. Klages},
\author[label11]{M. Kleifges},
\author[label5]{S. Kuhlmann},
\author[label4]{G. Mazzitelli},
\author[label3]{L. Nozka},
\author[label1]{A. Obermeier},
\author[label3]{M. Palatka},
\author[label6]{S. Petrera},
\author[label7]{P. Privitera},
\author[label3]{J. Ridky},
\author[label6]{V. Rizi},
\author[label7]{G. Rodriguez},
\author[label6]{F. Salamida},
\author[label3]{P. Schovanek},
\author[label5]{H. Spinka},
\author[label7]{E. Strazzeri},
\author[label12]{A. Ulrich},
\author[label5]{Z.M. Yusof},
\author[label8]{V. Vacek},
\author[label13]{P. Valente},
\author[label7]{V. Verzi},
\author[label10]{T. Waldenmaier},

\address[label2]{University of Chicago, Enrico Fermi Institute, 5640 S Ellis Ave., Chicago, IL, 60637, United States}
\address[label3]{Institute of Physics of the Academy of Sciences of the Czech Republic, Na Slovance 2, CZ-18221 Praha 8, Czech Republic}
\address[label4]{Laboratori Nazionali di Frascati dell'INFN, INFN, Sezione di Frascati, Via Enrico Fermi 40, Frascati, Rome 00044, Italy}
\address[label5]{Argonne National Laboratory, Argonne, IL, 60439, United States}
\address[label6]{Dipartimento di Fisica dell'Universit\`{a} de l'Aquila and INFN, Via Vetoio, I-67010 Coppito, Aquila, Italy}
\address[label7]{Dipartimento di Fisica dell'Universit\`{a} di Roma Tor Vergata and Sezione INFN, Via della Ricerca Scientifica, I-00133 Rma, Italy}
\address[label8]{Czech Technical University, Technicka 4, 16607 Praha 6, Czech Republic}
\address[label9]{Departmento de F\'{i}sica de Part\'{i}culas, Campus Sur, Universidad, E-15782, Santiago de Compostela, Spain}
\address[label1]{Universit\"at Karlsruhe (TH), Institut f\"ur Experimentelle Kernphysik (IEKP), Postfach 6980, D-76128 Karlsruhe, Germany}
\address[label10]{Forschungszentrum Karlsruhe, Institut f\"ur Kernphysik, Postfach 3640, D-76021 Karlsruhe, Germany}
\address[label11]{Forschungszentrum Karlsruhe, Institut f\"ur Prozessdatenverarbeitung und Elektronik, Postfach 3640, D-76021 Karlsruhe, Germany}
\address[label12]{Physik Department E12, Technische Universit\"at M\"unchen, James Franck Str. 1, D-85740 Garching, Germany}
\address[label13]{Sezione INFN di Roma 1, Ple. A. Moro 2, I-00185 Roma,Italy}
\corauth[cor1]{corresponding author}

\begin{abstract}
  One of the goals of the AIRFLY (AIR FLuorescence Yield) experiment
  is to measure the absolute fluorescence yield induced by electrons
  in air to better than 10\% precision.  We introduce a new technique
  for measurement of the absolute fluorescence yield of the
  \unit[337]{nm} line that has the advantage of reducing the
  systematic uncertainty due to the detector calibration. The principle is
  to compare the measured fluorescence yield to a well known process
  -- the \v Cerenkov emission. Preliminary measurements taken in
  the BFT (Beam Test Facility) in Frascati, Italy with 350 MeV electrons are
  presented. Beam tests in the Argonne Wakefield Accelerator at the
  Argonne National Laboratory, USA with 14 MeV electrons have also shown
  that this technique can be applied at lower energies. 
\end{abstract}

\begin{keyword}
Air fluorescence detection, Ultra high energy cosmic rays
\PACS 96.50.-S;96.50.sb;96.50.sd;32.50.+d;33.50.-j;33.50.Fa;34.50.Gb

\end{keyword}
\journal{5th Fluorescence Workshop, Madrid, 2007}
\end{frontmatter}
%\vskip 20pt
% main text
\section{Introduction}
\label{Sec:Intro}

The detection of ultra high energy ($\gtrsim 10^{18}$eV) cosmic rays
using nitrogen fluorescence emission induced by extensive air showers
(EAS) is a well established technique \cite{flyseye}. Atmospheric
nitrogen molecules, excited by EAS charged particles (mainly
$e^{\pm}$), emit fluorescence light in the $\approx$ 300--400 nm range.
The fluorescence detection of UHECR is based on the assumption that
the number of fluorescence photons of wavelength $\lambda$ emitted at
a given stage of a cosmic ray shower development, {\it i.e.} at a
given altitude $h$ in the atmosphere, is proportional to the energy
$E_{dep}^{shower}(h)$ deposited by the shower particles in the air
volume. Since a typical cosmic ray shower extends up
to about 15 km altitude, the fluorescence yield must be known over a
wide range of air pressure and temperature. Measurements by AIRFLY of the
fluorescence yield dependence on atmospheric parameters (pressure,
temperature and humidity), together with the spectral distribution between
 280 nm and 430 nm are presented in two separate contributions \cite{Andreas,Hal}.  

 It should be noted that  $E_{dep}^{shower}(h)$ is the sum
 of the energies deposited by EAS particles with a spectrum spanning
 from keV to GeV.  It is thus important to verify the proportionality
 of the fluorescence emission to the energy deposit over a wide range
 of electron energies. In \cite{EnergyScan}, the proportionality of the
fluorescence light to the energy deposit at a few \% level was tested
over the energy ranges 0.5 to 15 MeV, 50 to 420 MeV and 6 to 30 keV.
However, only relative measurements within each range were performed,
and absolute measurements of the fluorescence yield are in principle
needed to verify that the proportionality constant is the same in the
three measured energy ranges.

The absolute fluorescence yield is currently one of the main
systematics on the cosmic ray energy determination by EAS experiments
which employ the fluorescence technique. It is only known at the
level of \unit[15]\% and for a few electron energies \cite{others}. In this work, we will
report preliminary results of the measurements of the absolute
fluorescence yield of the most prominent line -- 2P(0,0) 337~nm by a
technique intended to keep the systematics below  \unit[10]\%.  The
data were taken in the BTF (Beam Test Facility) of the INFN Laboratori
Nazionali di Frascati, which can deliver 50--800 MeV electrons. Additionally,
  we have performed several tests at the the Argonne Wakefield Accelerator
(AWA), located at the Argonne National Laboratory, which can deliver
3--15 MeV electrons. The results presented here are preliminary
 and the intention of the authors is to show that the methodology 
 is appropriate to achieve accuracies below the \unit[10]\% level.

 The paper is organized as follows: in Section \ref{Sec:BTF} the technique
 proposed to measure the absolute fluorescence yield is presented and
 applied to the measurements in the BTF; in Section \ref{Sec:Argonne} recent
 measurements in the AWA are presented and the additional systematic
 uncertainties due to the smaller electron energies discussed; in Section
 \ref{Sec:Concl} we conclude and discuss future work.

\begin{figure*}[hbt]
\begin{equation}
\begin{split}
\underbrace{N_{fl}(337)}\ &=\quad \underline{\fly} \quad\times\ \ \underbrace{G_{fl}}\ \times\  
\underbrace{T\ \times\ \ Q(337)}\ \ \times\ \ 
\underbrace{N_{e^-}}\cr
\overbrace{N_{c}(337)}^{measured}\ &=\ \overbrace{\mind Y_{c}}^{known}\ \times\
\overbrace{G_{c}}^{MC}\ \times\ \
\overbrace{T\ \times\ Q(337)}^{\sim cancel}\ \times\overbrace{N_{e^-}}^{relative}\times\ \overbrace{R_{m}}^{measured}
\end{split}
\label{eq:cer/fl}
\end{equation}

\end{figure*}

\section{Absolute Fluorescence Yield Measurements at 350 MeV}
\label{Sec:BTF}

\subsection{Description of the method}

\begin{figure}[hbp]
  \centering
  \includegraphics[width=8.3cm]{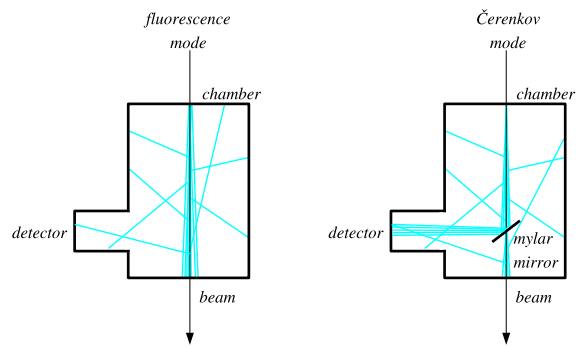}
  \caption{Setup for the measurement of the absolute
fluorescence yield. Remotely controlled mirror allows to switch between \v Cerenkov
and fluorescence modes without beam
interruption.}
  \label{fig:cere/fluo}
\end{figure}

AIRFLY uses a pressure chamber constructed of an aluminum tube with
various flanges welded to it for windows, gauges, gas inlet and
pump-out.  %The aluminum tube had an inner diameter of 201 mm, length
% 378 mm, and wall thickness 3 mm. The exit window of 0.1 mm thick aluminum
% was bolted to the one end. The entrance window of 0.5 mm thick beryllium, 35 x 55 mm,
% was diffusion bonded to a Conflat flange.
The electron beam passes through the axis of the chamber. A photon
detector, with a 337 nm interference filter in front, is placed in one
of the flanges perpendicular to the chamber axis. The measurements are
taken in two modes sketched in Fig. \ref{fig:cere/fluo}. In the
fluorescence mode, the isotropic fluorescence light produced by the
electrons in the field of view of the detector is recorded. In this
mode, contributions from other sources of light, like \v Cerenkov or
transition radiation, are negligible due to the non-isotropic emission
of such mechanisms. In the \v Cerenkov mode a thin mylar mirror at an
angle of $45^\circ$ is inserted remotely into the beam, redirecting
the \v Cerenkov light into the detector. In this mode, the \v Cerenkov
light fully dominates over fluorescence.

The absolute fluorescence yield is then determined using the ratio of
the signal measured in the fluorescence and in the \v Cerenkov configurations
from Equations \ref{eq:cer/fl}. 
The \v Cerenkov yield $ Y_{c}$ is known from the theory, the apparatus
geometrical factors $G_{fl}$ and $G_{c}$ are derived from
the full Geant4 simulation of the detector and take into account the probability
of a photon being emitted in each case and also the fact that \v Cerenkov light
is very directional and fluorescence is emitted isotropically . Relative number of
incident electrons is measured by monitoring devices. The filter
transmittance $T$ and the detector quantum efficiency
$Q(337)$ are identical in both configurations and therefore
cancel. The mylar mirror reflectivity $R_{m}$ was measured.

\subsection{Experimental setup and data analysis}

The BFT (Beam Test Facility) in Frascati is capable of delivering
electrons of energy 50 to 800~MeV and positrons of energy 50 to 550~MeV, with
intensities ranging from a single particle to 10$^{10}$ particles per bunch at a
repetition rate up to 50~Hz. The typical pulse duration is 10~ns.
The absolute fluorescence yield was measured at 350 MeV.

A hybrid photodiode (HPD) capable of single photoelectron counting was
used as the main photodetector. The photocathode was placed 202~mm
from any beam and the optical path was baffled to avoid any
reflections off the housing walls.

%\begin{figure}[hbp]
%  \centering
%  \includegraphics[width=8.cm]{fov.eps}
%  \caption{ Schematic view of the AIRFLY chamber in \v Cerenkov mode. The grey shaded
%region represents a GEANT4 simulation of the detector field of view.}
%  \label{fig:fov}
%\end{figure}

\begin{figure}[hbp]
  \centering
  \includegraphics[width=7.5cm]{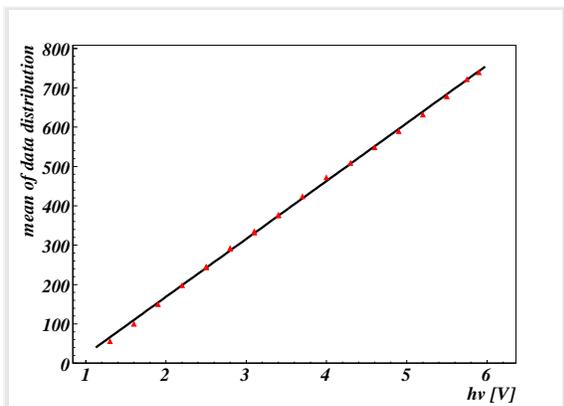}
  \caption{ Dependence of the mean value of the ADC signal on
 the HPD high voltage.}
  \label{fig:hpdlin}
\end{figure}

A 337~nm interference filter was placed in front of the HPD, the
aperture was limited to 40~mm at 60~mm perpendicular distance from the
beam to allow only up to 20$^\circ$ angle of incidence. A fast
scintillator 100 by 100~mm, 5~mm thick, was used to monitor the
beam intensity. The beam intensity was also monitored by NaI(Tl)
calorimeter with excellent single electron resolution, placed at the
end of the beam line. 

\begin{figure}[hbp]
  \centering
  \includegraphics[width=7.5cm]{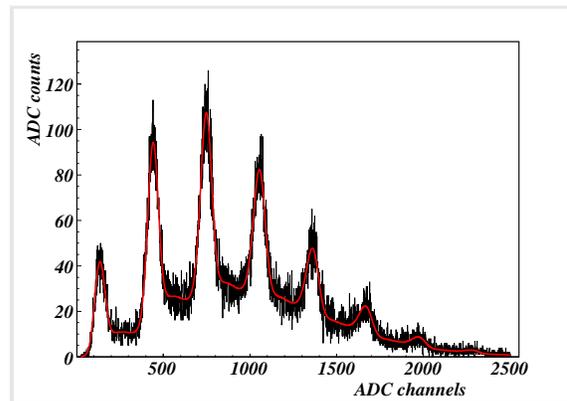}
  \caption{ An example of HPD signal (dots) fitted by the Simulated-annealing 
routine (continuous line) .}
  \label{fig:hpdfit}
\end{figure}

To improve the signal to noise ratio in the fluorescence mode, it was
necessary to maximize the dynamic range of the HPD by changing the
high voltage. The fluorescence data were taken at the highest possible
and the \v Cerenkov data at the lowest possible high voltage applied
to the HPD.  Linearity of the HPD response with respect to the high voltage 
was verified and is shown
in Fig.\ref{fig:hpdlin}. In the fluorescence mode, due to the small
number of photons, the analysis of the Frascati runs was done using
the single photoelectron signals from the HPD.  Intensity of the fluorescence
signal in number of 
photoelectrons was obtained using the model described in \cite{Tab97}, which
takes into account backscattering of the photoelectrons. 
The simulated-annealing fitting method \cite{GSL06}, \cite{Boh06} was used to
fit the model to the data. It is a Monte-Carlo minimization routine that has the
advantage of being able to escape from local minima.
  An example of the HPD signal fitted by the Simulated-annealing routine is
shown in Fig. \ref{fig:hpdfit}. 
The background was determined in the same way and subtracted.

In the \v Cerenkov mode, due to the
large number of photons, the average signal in ADC counts is
calculated, the background subtracted and converted into number of
photoelectrons. Consequently it was necessary to determine the
conversion factor from ADC counts to photoelectrons ($C_{c}$) as a
function of voltage. This was done by two methods: using LED with
the intensity adjusted so that individual photoelectrons are visible,
and using the \v Cerenkov signal from the beam with two different
intensities monitored by the scintillation palette.  Both methods gave
a consistent result within \unit[2]\%.

\begin{figure}[hbp]
  \centering
  \includegraphics[width=7.5cm]{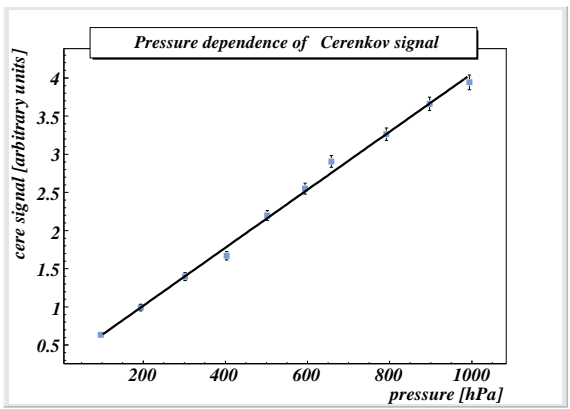}
  \caption{Dependence of the \v Cerenkov signal on pressure. A good linearity
 is observed, see text for more details.}
  \label{fig:cer-pres}
\end{figure}

The illumination of the photocathode is different in \v Cerenkov mode
(a light cone) and fluorescence mode (uniform illumination). The
photocathode uniformity is important to understand the detector
systematic uncertainties. To understand it, the \v Cerenkov signal
 was measured for different pressures. As the pressure decreases the \v Cerenkov
light-cone becomes narrower and also the multiple scattering of
electrons becomes smaller making the \v Cerenkov light spot to cover a
smaller part of the photocathode.  The measured dependence of the \v
Cerenkov signal on the pressure is shown in Fig. \ref{fig:cer-pres}.  A
nonlinearity for high pressures would indicate that some \v Cerenkov
photons are lost on the way to photocathode.  Also a nonuniformity of
the photocathode would spoil the linearity of this plot.  From the
good linearity of the \v Cerenkov pressure dependence
(see Fig. \ref{fig:cer-pres}) it is possible to deduce that the part of \v
Cerenkov light falling outside of the photocathode is negligible and
the nonuniformity, if any, should be limited to the very edges of the
photocathode.

\subsection{Monte Carlo Simulations and checks}

\begin{figure}[hbp]
  \centering
  \includegraphics[width=7.5cm]{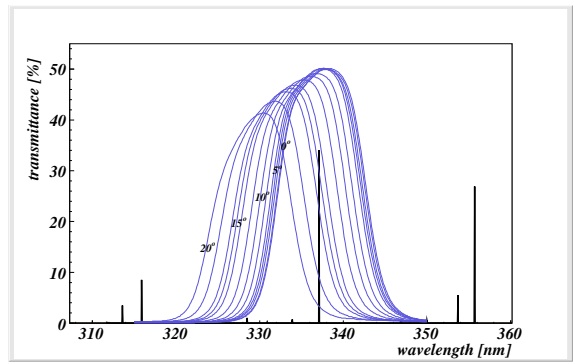}
  \caption{ Measured transmittance of the \unit[337.1]{nm} interference filter 
  at 13 different angles of incidence between $0^\circ$ and $20^\circ$ on top of
  the schematic nitrogen spectrum.}
  \label{fig:filtr}
\end{figure}

A full Monte Carlo simulation of the experimental setup was performed using
 Geant4 \cite{GEA04}.  In the simulation, the fluorescence yield was set to
\unit[19]{photons/MeV} deposited in a step sampled from the spectrum
of Bunner (280 -- 520 nm) \cite{Bun67}. This number leads to
$\unit[4.74]{photons/m/e^-}$ at \unit[350]{MeV} in $\unit[1]{m^3}$ of air
or $\unit[1.24]{photons/m/e^-}$ at \unit[337.1]{nm} line in the same
volume ($\unit[4.17]{photons/m/e^-}$ between \unit[300 - 400]{nm}) at the
pressure \unit[993]{hPa} and temperature $\unit[18]{^\circ C}$. The
\unit[337.1]{nm} line then forms \unit[26.2]\% of the total number of
photons. As the filter transmittance strongly depends on the incidence
angle (see Fig. \ref{fig:filtr}) and the distribution of incident angles in the fluorescence and
\v Cerenkov case differs significantly these effects have to be
included in the simulation of the geometrical factor. The default GEANT4
implementation of the \v Cerenkov process was used \cite{GEA04}.

\begin{figure}[hbp]
  \centering
  \includegraphics[width=7.8cm]{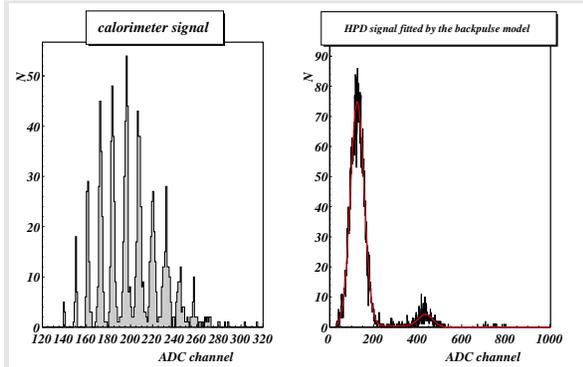}
  \caption{Calorimeter and the corresponding
  \v Cerenkov signal detected by HPD.}
  \label{fig:cercalo}
\end{figure}

To check our simulations, the number of \v Cerenkov photons detected
per \unit[350]{MeV} primary electron was experimentally measured and
compared with Monte Carlo results. The measurement was done in pure
nitrogen which has a slightly higher index of refraction than air.
The beam intensity was reduced to the point where it was possible to
see individual electrons in the calorimeter and also single
photoelectrons in the HPD. The HPD signal was analyzed by fitting the
backpulse model using the Simulated-annealing method to the data. An
example of data histogram from the calorimeter and corresponding
signal in HPD are shown in Fig. \ref{fig:cercalo}.  The background was
estimated at \unit[0.0021]{p.e./bunch}.  The calorimeter signal was
analyzed in the following way: number of hits in each individual peak
was multiplied by the corresponding number of electrons, summed up and
divided by the number of bunches.  Another approach is to fit a
Poisson distribution into the number of hits for each peak. Average of
three consecutive run yields $\unit[0.0158 \pm
0.0003]{photoelectrons/e^-}$.

This result can be then compared with the full Monte Carlo
simulation. Assuming the quantum efficiency reported by the
manufacturer (\unit[24.3]\% at 337 nm) and using the measured filter
transmittance interpolated between the angles and wavelengths, we
obtained an average number of \v Cerenkov photons detected per 350 MeV
primary electron of 0.0152, in agreement with the data.

\subsection{Preliminary results}
\label{subsec:preliminary}
The resulting fluorescence yield of the \unit[337]{nm} line is derived as
\begin{equation}
\fly =\  \underbrace{{N_{fl}(337)\over S_{c}(337)} \times
\ C_{c}}_{\mind R_{meas}}\ \times
 {N_{c}^{e^-}\over N_{fl}^{e^-}} \times {1\over R_{sim}}\times \mind Y_{i},
\end{equation}
where $N_{fl}(337)$ is the measured fluorescence signal in photoelectrons, 
$S_{c}(337)$ is the measured \v Cerenkov signal in ADC counts and
$C_{c}$ is the conversion factor from ADC counts to photoelectrons. 
The ratio ${N_{c}^{e^-}\over N_{fl}^{e^-}}$ is
measured by a fast scintillator. The ratio of fluorescence to 
\v Cerenkov signals obtained from simulations $R_{sim}= 1.330\times 10^{-3}$ is
proportional to the $\mind Y_{i}$ (which enters the simulation).
The measured ratio yields $R_{meas}= 1.10\times 10^{-3}$. 

The preliminary absolute fluorescence yield derived at \unit[993]{hPa} and
$\unit[18]{^\circ C}$ is
\begin{equation}
\unit[19]{ph/MeV} \times 0.262 \times 1.10/1.330 
 = \unit[4.12]{ph/MeV}\,.
\end{equation}
The quantity \unit[19]{ph/MeV} and the relative spectrum is contained
also in the simulated ratio $R_{sim}$ and the two occurrences cancel,
therefore the resulting fluorescence yield does not depend on the choice of initial
values. While our result is still preliminary, it compares well with the value of \unit[4.32]{ph/MeV} deposited  at sea level quoted by Bunner \cite{Bun67}. Also, it is within 20\% of the fluorescence yield that may be derived from other more recent measurements \cite{others}. 
This agreement strengthens our confidence in the validity of the method.   

%The ratio of the fluorescence yield in nitrogen and air measured for
%the \unit[337]{nm} line was found to be
%\begin{equation}
%{\fly(337)_{N_2}\over \fly(337)_{Air}} = 7.75. 
%\end{equation}

\subsection{Discussion of systematic errors}

Many tests and simulations were done in order to understand the
systematic uncertainties of the absolute measurement.  The measured
background was very small -- about $\unit[(0.006\pm0.001)]{p.e./bunch}$
and amounted to about \unit[10]\% of the fluorescence signal. In the
\v Cerenkov case the background is negligible so the uncertainty
introduced by the background subtraction is \unit[2]\% (statistical uncertainty).  The beam
intensity normalization is a relative quantity. The statistical error
is negligible. The uncertainty of \unit[1]\% assigned to this aspect accounts
for the possible nonuniformity in the plastic scintillator response.

The HPD fitting method uncertainty stems from the fact that the merit
function has a broad minimum causing that a slight change in the
fitted parameters will lead to a slightly different number of
photoelectrons but the quality of the fit will stay the same. This
effect was estimated to amount to \unit[3]\%.  HPD calibration and
filter transmittance were already discussed previously. 
The mylar mirror reflectivity at the wavelength range needed was measured to be
\unit[$(84\pm 1)$]\%. 
Systematic effect caused by a slight misalignment of the detector
components was studied in simulations.  The
photocathode nonuniformity is currently under study. The photocathode
coverage differs significantly between the fluorescence and \v
Cerenkov case so any radial nonuniformity could influence the ratio in
a substantial way.  Therefore the largest systematic error was
assigned to it.  Current estimates of the systematic uncertainties taken into account are summarized in table \ref{tab:syster}.

\begin{table}
\begin{tabular}{cc}
\hline
    background subtraction  & $2\%$ \\
    beam intensity normalization & $1\%$  \\
    beam position and spotsize  & $1\%$ \\
    geometry (misalignment)  & $4\%$ \\
    HPD fitting method  & $3\%$ \\
    HPD calibration (ADC/p.e.)  & $2.3\%$ \\
    simulation (model)  & $2\% $  \\
    $45^\circ$ mirror reflectivity & $1.2\%$ \\
    \unit[337]{nm} filter transmittance & $2\%$ \\
    photocathode uniformity and angular dependence& $5\%$ \\
\hline
\end{tabular}
  \caption{Systematic errors of the absolute fluorescence yield
    measurement.}
  \label{tab:syster}
\end{table}

Contamination by the \unit[333.9]{nm} line is below \unit[1]\%. Also
the contribution of transition radiation from the mirror was found to
be negligible.
 
Statistical uncertainty amounts to \unit[1.5]\%. 

%As the contributing uncertainties
%are independent they were added in quadrature and the resulting error
%to be assigned to the absolute measurement is \unit[8.5]\%.

The systematic uncertainties in table \ref{tab:syster} are still preliminary. In particular, additional work is needed to understand the photocathode nonuniformity, and dedicated measurements are foreseen. Also, the AIRFLY apparatus was moved to Argonne after this first measurement, which did not allow for additional tests and crosschecks of the result. New measurements with an improved apparatus are foreseen.
 
Nevertheless, this estimate shows that a systematic uncertainty  $<10\%$ on the absolute yield can be achieved with the experimental method proposed in this paper.  

\section{Absolute Fluorescence Yield Measurements at 15 MeV}
\label{Sec:Argonne}
As it was mentioned in Section \ref{Sec:Intro},  absolute fluorescence yield 
 measurements at 15 MeV would allow us to verify the proportionality
 of fluorescence yield and energy deposit in the energy range 1--400 MeV.
Additionally, it will confirm the results presented in the previous section
 with a measurement that suffers from different systematic errors.
 In this line, measurements were performed at the Argonne Wakefield
 Accelerator (AWA), located at the Argonne National Laboratory. The
 measurement principle is the same as outlined before, i.e.
 the absolute fluorescence yield is obtained from the measured ratios
 of the signal in the \v Cerenkov and fluorescence mode (Fig.
 \ref{fig:cere/fluo}). 

 The LINAC was able to deliver electrons in the energy range 3--15 MeV. It was 
 operated at 5 Hz, with bunches of maximum charge of 1
 nC and length 15 ps (FWHM) and typical energy spread of $\pm$ 0.3 MeV
 at 14 MeV.  The beam spot size was typically 5 mm diameter, with
 negligible beam motion. The beam intensity was monitored with an
 integrating current transformer (ICT), placed directly before the beam
 exit flange. The signal from the ICT was integrated, digitized, and
 recorded for each beam bunch.  The \v Cerenkov/fluorescence light was detected by a
 photomultiplier tube (PMT Hamamatsu H7195 model) with a narrow band 337
 nm filter, located about 80 cm away from the beam axis. A shutter
 installed in front of the PMT allowed measurements of background. The
 PMT was surrounded by considerable lead shielding to reduce
 beam-related backgrounds.
 
 The electrons in this energy range are below the \v Cerenkov
 threshold in air, and, for that reason, we used in the \v Cerenkov
 mode gases with larger refraction indexes (Freon-12 and SF$_6$). The
 high voltage settings for the PMT were the same in the \v Cerenkov
 and fluorescence mode.  Due to the large number of photons in the \v
 Cerenkov mode ($\sim$ 1000 times larger than in the fluorescence
 mode) and to avoid the PMT saturation, an attenuation filter is
 placed in front of the PMT during the \v Cerenkov measurements.

 The LINAC was operated in a mode allowing the bunch charge to
 fluctuate over a wide range. The correlation of the PMT and ICT
 signals, which showed a linear relation, was fitted and the slope
 $S_{meas}$ was taken as an estimator of the fluorescence signal. The
 same procedure was applied with the shutter closed to estimate the
 background, which was subtracted.

\begin{figure}[hbp]
  \centering
  \includegraphics[width=7.8cm]{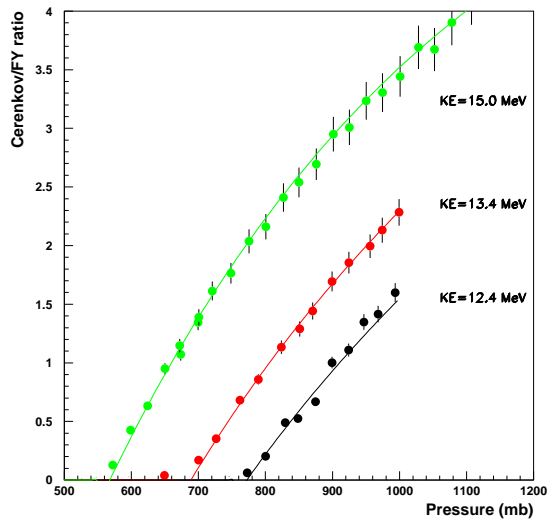}
  \caption{The measured ratios of the 
 \v Cerenkov to fluorescence signal as a function of pressure for three
 electron energies ( 12.4, 13.4 and 15 MeV). The \v Cerenkov gas used was Freon-12. }
  \label{fig:freonfeb06}
\end{figure}

 Fig. \ref{fig:freonfeb06} shows the measured ratios of the \v
 Cerenkov to fluorescence signal as a function of pressure. Systematic uncertainties on the measured ratios were estimated from run to run fluctuations. The
 fluorescence signal was measured in air at a fixed value of pressure
 (\unit[800]{hPa}), temperature (ambient) and for an electron energy
 of 13.4 MeV. The \v Cerenkov signal was taken in Freon-12 for a range
 of pressures and 3 electron energies ( 12.4, 13.4 and 15 MeV). The
 data were taken during the test beam on February 2006. It should be noted that the
 threshold for \v Cerenkov production is given by $n~\beta$ being $n$
 the refraction index of the gas and $\beta$ the speed of the
 electron. In this work the refraction index of Freon-12 was
 treated as an unknown and compared with the values quoted in the
 literature.

 Full Monte Carlo simulations of the detector setup were performed
 using Geant4. The predicted ratios of \v Cerenkov and fluorescence
 signal (at the energy and pressure indicated above) were
 parameterized as a function of electron energy, pressure and
 refraction index of the \v Cerenkov gas. It should be noted that this
 parameterization should be the \v Cerenkov formula with a
 normalization given by simulations at a specific energy, pressure and
 refraction index.  However, we have found that the Coulomb scattering
 of electrons travelling through the beam exit window, chamber
 entrance window and the gas, distorts the \v Cerenkov cone acting as
 a diffuser.  Consequently, most of the \v Cerenkov cone is not
 contained in the photocathode and only \unit[$\sim$25]\% (for p=800 mb,
 E=13.4 MeV and $n$=1.00108) of the light is detected in most of the
 cases.  Therefore, a more complicated parameterization is performed.

 A maximum likelihood method was then applied to fit all the data in
 Fig. \ref{fig:freonfeb06}.  The free parameters fitted were: the
 refraction index of the gas at 1000 mb ($n$), the absolute
 fluorescence yield for the 337 nm line, and three offsets for the
 electron energy. In the likelihood function we also include three
 terms to account for the fact that the nominal electron energies have
 an uncertainty of 0.3 MeV. Results of the fit are the solid lines in
 Fig. \ref{fig:freonfeb06}.
%absolute fluorescence yield 5.42$\pm$0.12 photons per MeV.  
The energy offsets were all below 0.3 MeV  and the refraction index $n$=1.00101 $\pm$0.00001 is reasonably compatible with
 what we found in the literature ($n$=1.00108). The absolute fluorescence yield
 was within $\approx 25\%$ of the value obtained at 350 MeV (see Section \ref{subsec:preliminary}). 

Two more test beams were performed in December 06 and February 07, 
 using different experimental setups, optical elements and two different
 gases (Freon-12 and $SF_6$). The data sets are not as complete as the one
 presented before, but we applied the same analysis and we found the
 same absolute fluorescence yield within \unit[5]\%.
 
 The systematics of this measurement are currently under
 investigation.  Preliminary studies indicate that variations
 of 1 degree in the beam angle can lead to variations of \unit[10]\% in the
 absolute fluorescence yield.  The \v Cerenkov light distribution in
 the photocathode is very uniform due to the Coulomb scattering
 effect, and therefore the systematics due to non-uniformities of the
 photocathode should be much less important than in the Frascati
 measurement. Contribution from transition radiation was found to be negligible.
The most important systematic is the modelling of the Coulomb scattering in the full
 Monte-Carlo simulations.  Coulomb scattering is a well know
 microscopic process, but its modelling in the Monte Carlo is not done
 in a microscopic way, therefore an algorithm has to be adopted.
 We used the algorithm implemented in Geant4 \cite{Urban}, we checked
 that the angular distribution of electrons close to the mylar mirror
 fits rather well the analytical formula for multiple scattering 
from Moliere Theory in
 \cite{PDG}.  We have also varied the multiple scattering according to
 the errors of $\theta_{rms}$ quoted in \cite{PDG} and observed
 variations of \unit[$\sim$ 5]\% in the number of \v Cerenkov photons. However,
 and since this measurement heavily relies on the modelling of this effect,
 an experimental proof of the goodness of this modelling is necessary.

%In summary, the absolute fluorescence yield obtained for the 337 nm line
% at 15 MeV is \unit[22]\% higher than the value quoted in the previous
% section at 350 MeV. The statistical error including background subtraction
% is smaller than \unit[2]\%. 

The assessment of the systematic errors is an undergoing
 process, and we estimate that the current uncertainty is not smaller than 15\%. It is clear that at 15 MeV we are more model dependent due to
  multiple scattering of electrons. The AIRFLY chamber was designed to perform the absolute yield measurement at GeV energies, where multiple scattering is small. We are currently studying experimental
 setups to reduce the effect of multiple scattering in this measurement, as well
 as ways to verify the Monte Carlo modelling of this effect.

\section{Outlook}
\label{Sec:Concl}

A novel technique to measure the absolute fluorescence yield of the
most prominent line -- 2P(0,0) 337~nm has been presented.

Preliminary measurements performed at the BTF facility in Frascati with 350 MeV electrons showed that a systematic uncertainty  below the \unit[10]\%
level is within reach.
 
We also investigated the feasibility of the absolute yield measurement with the
same technique at lower energies. We performed a series of beam tests in the
Argonne Wakefield Accelerator with 12 to 15 MeV electrons. The systematic
uncertainty of these measurements were found to be at least \unit[15]\% due to
the increased importance of multiple scattering at these lower energies. We are confident that a new experimental setup and dedicated measurements to verify the Monte
Carlo modelling of multiple scattering will eventually reduce the systematic uncertainty to the \unit[10]\% level.

AIRFLY has now completed the measurements of the air fluorescence spectrum dependence on atmospheric parameters. The pressure dependence has been published \cite{airflyAP}. The analysis of the temperature and humidity dependence is advanced, and results on a selected set of lines has been presented at this Workshop \cite{Hal}. AIRFLY will now focus on the measurement of the absolute yield, and measurements with an improved apparatus, which takes advantage of the experience gained with the preliminary tests reported in this paper, are foreseen both at GeV and MeV energies.

% \appendix

% \section{}
% \label{}

%\begin{thebibliography}{00}


% notes:

% \bibitem{label2}
% If there is a note, it should come last:
% \bibitem{label3} \note
% \end{subbibitems}

%\bibitem{}

%\end{thebibliography}

\section{Acknowledgments}
We thank the staff of Argonne National Laboratory for their support.
This work was also supported by the grant of MSMT CR LA 08016 and 1M06002 and
ASCR grants AV0Z10100502 and AV0Z10100522.
A.\ Obermeier and J.\ R.\ H\"orandel acknowledge the support of
VIHKOS, which made the participation at the measurement campaigns
possible.

\bibliographystyle{elsart-num}
\bibliography{biblio}

\end{document}